\documentclass[10pt]{iopart} % for two column
\usepackage{iopams}

\usepackage{graphicx}% Include figure files
\usepackage{dcolumn}% Align table columns on decimal point
\usepackage{bm}% bold math
\usepackage{color} % 色つけ用として
\usepackage{ulem} % 打ち消し線を使うために

\usepackage{textcomp} % textsymbol

\begin{document}

\title{Chirality-selected crystal growth and spin polarization over centimeters of transition metal disilicide crystals}

%\author{Content \& Services Team}
\author{Yusuke Kousaka$^{1*}$, Taisei Sayo$^1$, Satoshi Iwasaki$^2$\footnote{Present address: Novel Crystal Technology, Inc., Saitama, 350-1328, Japan}, Ryo Saki$^1$, Chiho Shimada$^1$, Hiroaki Shishido$^1$, Yoshihiko Togawa$^1$}
\address{$^1$ Department of Physics and Electronics, Osaka Metropolitan University, Sakai, Osaka 599-8531, Japan}
\address{$^2$ Research Institute for Interdisciplinary Science, Okayama University, Okayama, Okayama 700-8530, Japan}

\ead{koyu@omu.ac.jp}

\vspace{10pt}
\begin{indented}
\item[]September 2022
\end{indented}

\begin{abstract}
We performed a chirality-controlled crystal growth of transition metal disilicide NbSi$_{2}$ and TaSi$_{2}$ by using a laser-diode-heated floating zone (LDFZ) method.
The crystal chirality was evaluated in the crystals of centimeters in length by performing single crystal X-ray diffraction as well as probing a spin polarization originating from chirality-induced spin selectivity (CISS) effect.
The crystals of right-handed NbSi$_{2}$ and of left-handed TaSi$_{2}$ were obtained in the conventional LDFZ crystal growth, while the left-handed NbSi$_{2}$ and right-handed TaSi$_{2}$ crystals were grown by the LDFZ method with the composition-gradient feed rods.
The spin polarization via the CISS was observed over centimeters in the NbSi$_{2}$ single crystals and the sign of the CISS signals was dependent on the chirality of crystals.
The correlation between the crystal chirality and CISS signals indicates that the CISS measurements work as a non-destructive method for chirality determination even in centimeter-long specimens.
\end{abstract}

%\maketitle
\ioptwocol

\section{Introduction}

Chirality, representing right- or left-handedness, generates attractive physical properties in material science.
The importance of chirality has been recognized by a relationship between chiral structures and chirality-induced phenomena,
as exemplified by natural optical activity in optical science \cite{Arago1811,Biot1812,Pasteur1,Pasteur2} and chiral helimagnetic ordering,
triggered by an antisymmetric exchange (Dzyaloshinskii-Moriya) interaction \cite{Dzyaloshinskii1958,Moriya1960},
in chiral magnetism \cite{TSi2,Ishida1985,Kishine2015,Togawa2016}.
Indeed, the sign of Dzyaloshinskii-Moriya interaction depends on the handedness of a crystalline structure of chiral helimagnetic materials \cite{Togawa2012}.
In this connection, the importance of synthesizing inorganic enantiopure crystals has been emphasized in material science \cite{Kousaka2017}, which will advance our understanding of a wide range of chirality-induced phenomena in inorganic chiral materials such as magnetochiral dichroism (MChD) \cite{Kibayashi2014,Okumura2015,Nakagawa2017},
nonreciprocal electrical transport \cite{Rikken2001,Yokouchi2017,Aoki2019} and chirality-induced spin selectivity (CISS) effect \cite{Inui2020,Shiota2021,Shishido2021}.

The CISS effect has attracted much attention since a controllability of spin polarization is given in nonmagnetic materials with chiral structures.
The amplitude of the spin polarization becomes as large as that for ferromagnetic materials, while its sign is dependent on the chirality of materials.
It was first observed in chiral molecules \cite{Gohler2011,Xie2011}, and quite recently found in chiral inorganic crystals \cite{Inui2020,Shiota2021,Shishido2021}.
In the latter case, when a charge current flows through the crystal, 
the spin-polarized state appears as a linear response of the current-voltage characteristics at room temperature without magnetic field.
The sign of the linear slope is connected to the crystal handedness. In this respect, the CISS signal has a great potential to probe the crystal chirality.
Furthermore, the spin polarization emerges in a length scale over micrometers \cite{Inui2020,Shiota2021} and even over millimeters \cite{Shishido2021} in nonlocal measurements with inorganic chiral crystals.

Note that it is still difficult to evaluate and control the chirality of crystal structures in inorganic materials.
The crystal chirality is generally determined by an absolute structure analysis using single crystal X-ray diffraction measurements.
One of the disadvantages of the X-ray method is that the sample size is limited to a scale of sub-millimeter or smaller because the structure analysis is significantly affected by a large X-ray absorption.
When the size of specimens is in a millimeter scale or larger, such crystals need to be smashed into small portions of sub-millimeter.
In this respect, the single crystal X-ray diffraction measurements correspond to a destructive method for chirality evaluation and probe the chirality in a part of the specimens.

As for controlling the chirality, inorganic chiral compounds normally crystallize racemic twinned crystals, which contain right- and left-handed chirality domains.
One possible method of obtaining a homochiral crystal with the selected handedness is a spontaneous crystallization with stirring solution \cite{Kondepudi1990}.
This technique is applied to a synthesis of water-soluble compounds.
Indeed, the mono-chiral crystals are successfully obtained in NaClO$_3$ \cite{Kondepudi1990} and CsCuCl$_3$ \cite{Kousaka2017,Kousaka2014}, leading to the observation of enhanced MChD signals \cite{Nakagawa2017}. 
Obviously, these studies indicate that the enantiopure crystals play a key role in inducing nontrivial physical properties coupled with the handedness of the crystals.

Another example of chirality-controlled synthesis is found in intermetallic monosilicide compounds $T$Si ($T:$ a transition metal) with a B20 type chiral cubic crystal structure.
$T$Si is known as one of the representative chiral magnetic compounds, which exhibit a formation of chiral helimagnetic structure \cite{TSi2,Ishida1985,TSi3} and Skyrmion lattice \cite{Kadowaki1982,Bogdanov1989,Muhlbauer2009,Neubauer2009,Yu2010}.

It is known that single crystals of $T$Si are obtained with the single handedness by use of conventional crystal growth techniques and the crystal chirality depends on the element of $T$ site \cite{TSi2,TSi3,TSi4,Siegfried2015}.
The right-handed crystals are stabilized when $T =$ Fe, while the left-handed ones are grown when $T =$ Mn and Co.
Note that a seed crystal is used in a chirality-selected crystal growth.
Interestingly, the handedness of the grown crystal inherits from the one of the seed crystal. 
Namely, when using a seed crystal with the $T$ element appropriate for the desired handedness, the single crystal with the selected handedness is obtained \cite{TSi5,Kousaka2022,Bauer2022}.
For instance, a right-handed CoSi can be grown by using a right-handed FeSi as a seed crystal.
Such ideas of the chirality-controlled crystal growth were indeed demonstrated in a Czochralski method \cite{TSi5} and in a floating zone (FZ) method \cite{Kousaka2022,Bauer2022}.

Intermetallic disillicide compounds $T$Si$_{2}$ with a C40 type chiral hexagonal crystal structure are the next target materials for the chirality-controlled crytal growth.
Recently, $T$Si$_{2}$ is found to show the CISS effect \cite{Shiota2021,Shishido2021}.
The direction of the spin polarization in the right-handed NbSi$_{2}$ is opposite to that in the left-handed TaSi$_{2}$ \cite{Shiota2021}.
These crystals exhibit the spin polarization in a length scale over millimeters \cite{Shishido2021}.

The chiral crystals of centimeters in length are necessary in order to investigate the length scale over which the spin polarization is kept in the chiral crystal.
Moreover, it would be interesting to examine the spin polarization between the same compound $T$Si$_{2}$ of different chirality (the same $T$ disillicide with right-handed and left-handed structures). These studies may reveal a correlation between the crystal chirality and spin polarization via the CISS effect.

In this study, we demonstrate the chirality-controlled crystal growth of NbSi$_{2}$ and TaSi$_{2}$ by using a laser-diode-heated floating zone (LDFZ) method. 
With the present technique, we obtained the chirality-selected crystals of NbSi$_{2}$ and TaSi$_{2}$ over centimeters long.
The CISS experiments with NbSi$_{2}$ of the selected chirality unveiled a correlation between the crystal chirality and CISS signals.
Our study exhibits that the CISS measurements work as a non-destructive technique for the chirality evaluation even in centimeter-sized specimens.

\section{Crystal structure and chirality of disillicides}

\begin{figure}[tb]
\begin{center}
\includegraphics[width=8cm]{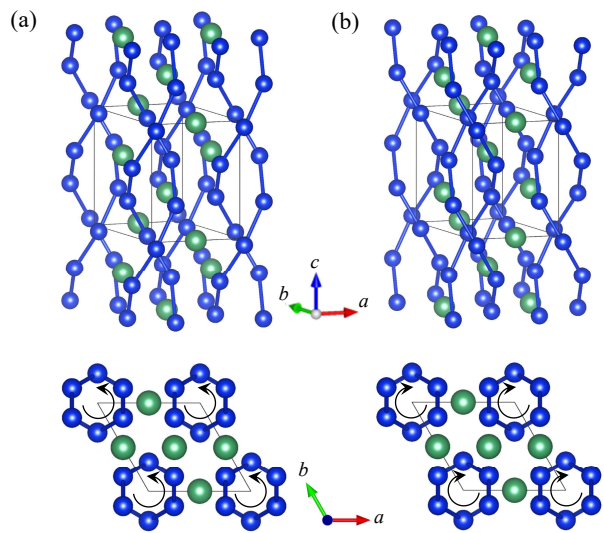}
\end{center}
\caption{Crystal structures of (a) right-handed and (b) left-handed $T$Si$_{2}$ ($T:$ a transition metal).
The lower schematics show the (001) projection of each structure.
Large green and small blue balls represent $T$ and Si atoms, respectively.
The black arrows are given as an eye guide to recognize the sense of screw alignments of the Si atoms.}
\label{f-TSi-str}
\end{figure}

The disillicides $T$Si$_{2}$ belong to the space group $P6_{2}22$ and $P6_{4}22$, which forms the right- and left-handed crystal structures, respectively, as shown in Fig.~\ref{f-TSi-str}. $T$ and Si atoms occupy Wyckoff $3c$ and $6i$ positions as expressed by $(1/2,0,0)$ and $(x,2x,0)$, respectively. 
The parameter $x$ for Si slightly deviates from a value of $1/6$.
Accurately speaking, the $T$ and Si atoms can also locate at Wyckoff $3d$ and $6j$ positions, respectively.
The difference between a pair of the $3c$ and $6i$ positions and that of the $3d$ and $6j$ positions is given by a displacement vector of $(0,0,1/2)$.
When the crystal structures are viewed along the $c$-axis, the right-handed double helices of Si atoms are found in the right-handed crystal and vice versa in the left-handed crystal.

The crystal chirality of the disillicide compounds is examined by an absolute structure analysis using single crystal X-ray diffraction, clarifying that the crystal chirality of $T$Si$_2$ depends on the elements of $T$ site \cite{Sakamoto2005, Onuki2014}.
The right-handed crystals are stabilized when $T =$ Nb, while the left-handed ones are grown when $T =$ V, Cr, and Ta.

\begin{figure*}[tb]
\begin{center}
\includegraphics[width=15.0cm]{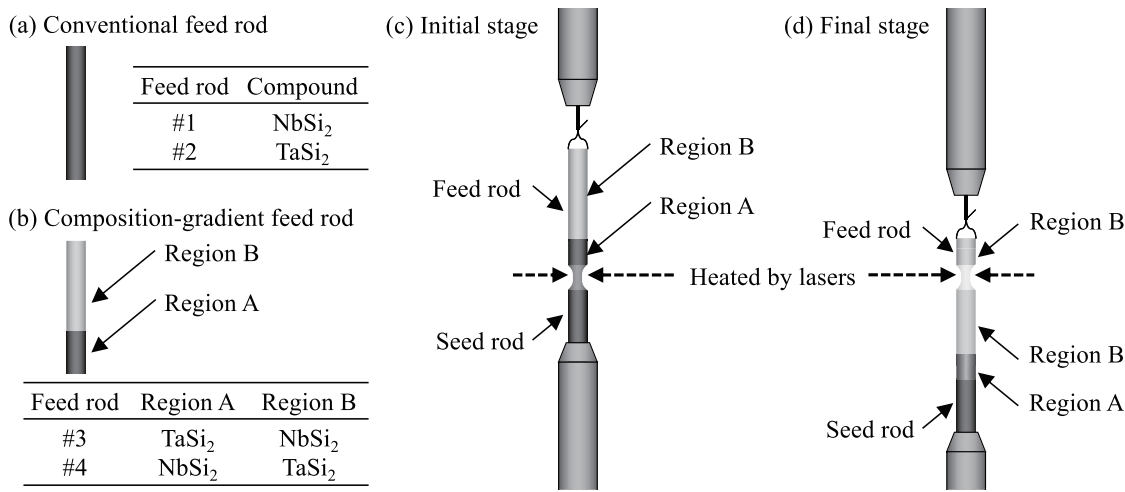}
\end{center}
\caption{
Schematic drawings of a crystal growth using a laser-diode-heated floating zone (LDFZ) method.
(a) A conventional LDFZ method using a feed rod made of the single compound.
(b) A LDFZ method using a composition-gradient feed rod, which consists of two regions with different compounds.
A combination of the $T$Si$_{2}$ compounds used in the present study is listed.
(c) and (d) The procedure of the LDFZ crystal growth.
The crystal growth starts with the region A, transfers into the region B, and finishes with the edge of the feed rod. As a result, the chirality of the region A is inherited into that of the region B. As for the composition-gradient feed rod (\#3) shown in Fig. \ref{f-FZ}(b), the chirality of TaSi$_{2}$ in the region B of the grown rod is determined by that of NbSi$_{2}$ (left-handed) in the region A, even though TaSi$_{2}$ prefers right-handed chirality in the conventional crystal growth. Various combinations of C40 materials in the feed rod enables the control of chirality of the grown crystals.
}
\label{f-FZ}
\end{figure*}

\begin{figure}[tb]
\begin{center}
\includegraphics[width=8cm]{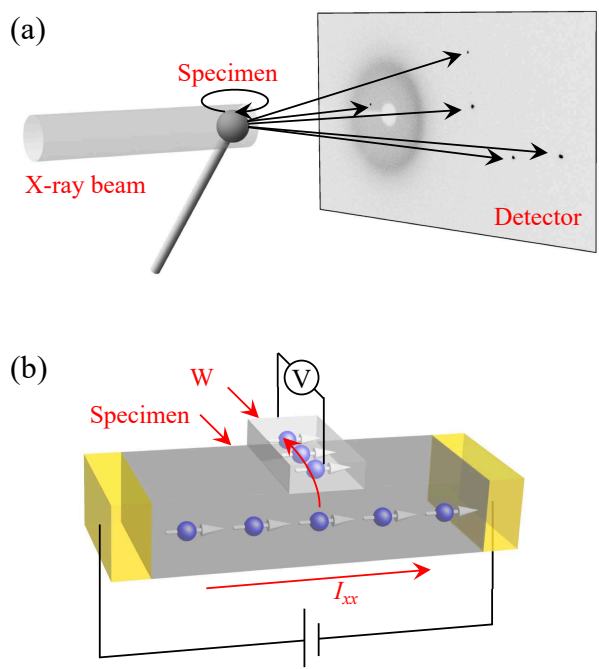}
\end{center}
\caption{
Schematic drawings of probing the chirality of materials using (a) a single crystal X-ray diffraction method and (b) electrical transport measurements via the CISS effect.
Note that the single crystal X-ray diffraction experiments were carried out by rotating the specimen with $0.5^\circ$ oscillation per frame. Over one thousand frames were collected for a specimen in order to observe thousands of Bragg reflection spots.}
\label{f-chiral-probe}
\end{figure}

\section{Experimental methods}

Single crystals of $T$Si$_{2}$ are obtained by the LDFZ method, as shown in Fig.~\ref{f-FZ}.
The FZ method is a crystal growth technique to obtain rod-shaped single crystals with several millimeters in diameter and several centimeters in length.
Generally speaking, the feed and seed rods for the crystal growth consist of the same compound.
In this study, the feed rods contain only NbSi$_{2}$ or TaSi$_{2}$ for the conventional FZ crystal growth, as shown in Fig.~\ref{f-FZ}(a).
A molten zone is formed by connecting the edges of both the rods that are melted by optical radiation sources.
In the growth process, the rods are pulled down through the optical radiated area. 
When the melting part moves out of the radiation, a single crystal is grown at the boundary between the melting and solidifying regions.

However, in some compounds, it is difficult to obtain single crystals via the conventional FZ crystal growth because of a very slow growth rate and/or instability of the molten zone.
Such problems can be solved by preparing a composition-gradient feed rod in the FZ crystal growth \cite{Ikeuchi2003, Enoki2010, Enoki2013}.
This advanced technique was successfully demonstrated for a growth of superconducting copper oxide compounds La$_{2-x}$Sr$_{x}$CuO$_{4}$ (La214) \cite{Ikeuchi2003} and Bi$_{2+x}$Sr$_{2-x}$CuO$_{6}$ \cite{Enoki2010, Enoki2013}.
In the case of La214, the composition-gradient feed rod had a step-like Sr-concentration gradient along the rod. Eventually, the obtained single crystal had an alternation of Sr concentration along the grown direction of the crystal.

Similar technique can be used in a synthesis of chiral inorganic compounds $T$Si.
The LDFZ method using the composition-gradient feed rod of $T$Si sustained the chirality inherited from the seed rod. The grown MnSi and CoSi crystals exhibited the chiral structures, which had the chirality opposite to that obtained by the conventional LDFZ method \cite{Kousaka2022}.

In this study, we performed the crystal growth of $T$Si$_{2}$ by the LDFZ using a composition-gradient feed rod consisting of TaSi$_{2}$ and NbSi$_{2}$, as shown in Fig.~\ref{f-FZ}(b).
To prepare the feed and seed rods, the rod-shaped polycrystalline samples of $T$Si$_{2}$ were synthesized by solid-phase reaction.
A stoichiometric mixture of $T$ and Si in the molar ratio of 1 : 2 was heated at $1100 ^\circ$C in a vacuumed quartz tube for two days.
Single crystals of $T$Si$_{2}$ were grown by the LDFZ furnace (equipped with five 200 W lasers, Crystal Systems Corporation).
The crystal growth was performed in an Ar atmosphere with a flowing rate of 3 $l$$/$min.
To obtain single crystals, traveling speeds of the feed and seed rods were set to be the same, and the maximum speed was 10 mm/hour.
To make a stable molten zone, the feed and seed rods were rotated in the opposite directions at 4 and 5 rpm, respectively.

The crystallographic orientation of the rod-shaped crystals was evaluated by a Laue X-ray backscattering method
using a tungsten X-ray tube and an imaging plate (IPX-LC, IPX Corporation).
The crystal chirality was probed by the absolute structure analysis using a single crystal X-ray diffractometer using a molybdenum X-ray tube (XtaLAB mini II, Rigaku Corporation), as shown in Fig.~\ref{f-chiral-probe}.
In addition, the CISS signals were obtained using electrical transport measurements via an inverse spin Hall effect.

The single crystal X-ray diffraction experiments were performed by rotating the specimen with $0.5^\circ$ oscillation per frame in the $\omega$-axis.
More than one thousand frames were collected to deduct integrated intensities of thousands of the Bragg reflection spots,
which are assigned to the Miller indices $(hkl)$ in a reciprocal hemisphere.
As shown in Fig.~\ref{f-chiral-probe}(a), the specimen size must be smaller than the beam diameter and thus is limited to the product $\mu t$, where $\mu$ and $t$ are X-ray absorption coefficient and path length, respectively.
Resultantly, the absolute structure analysis is applicable only to the specimens of a sub-millimeter scale or smaller one.

Regarding the CISS effect, the direction of spin-polarization depends on the crystal chirality of materials.
In inorganic chiral crystals, the spin-polarization has been probed by an inverse spin Hall effect, where a tungsten (W) or platinum (Pt) electrode, made on the rectangular-shaped specimen, is used for the spin detection, as shown in Fig.~\ref{f-chiral-probe}(b). 
There is no limitation on sample size in the CISS measurements. Namely, it ranges from micrometers or millimeters in length \cite{Inui2020,Shiota2021,Shishido2021}.
When a longitudinal electrical current $I_{xx}$ is applied to the $c$-axis of chiral crystal, the spin-polarization occurs in the crystal because of the CISS effect.
The electrode detects the difference of the spin-dependent chemical potential difference between the specimen and the electrode,
and converts the spin polarization into the transverse voltage $V_{xy}$ via the inverse spin Hall effect.
Therefore, the crystal chirality of the specimen can be determined by observing the sign of the slope of the transverse voltage $V_{xy}$.

\section{Experimental results}

\subsection{Chirality-controlled crystal growth}

First, the crystal growth was carried out by the conventional LDFZ method.
In this case, the seed and feed rods were made of the single compound of C40 materials, as shown in Fig.~\ref{f-FZ}(a).
The rod-shaped crystals of NbSi$_{2}$ (\#1) and TaSi$_{2}$ (\#2) were grown by the LDFZ method.

Laue X-ray backscattering experiments showed that the crystallographic $c$-axis of the NbSi$_{2}$ and TaSi$_{2}$ crystals oriented in the growth direction of the obtained rod with a slight deviation of a few degrees.
For chirality determination, a part of the edge of the crystal rod was smashed into some submillimeter pieces and these pieces were evaluated by means of single crystal X-ray diffraction.
The examined pieces of NbSi$_{2}$ and TaSi$_{2}$ crystals exhibited tiny mosaicity and were available for the absolute structure analysis.
With applying an empirical absorption correction to the observed intensity, the absolute structures were solved by the direct method and refined using the SHELXL software package \cite{Shelx}.

In the case of non-centrosymmetric crystal structures, the intensities of a pair of reflections at $(h k l)$ and $(\bar{h} \bar{k} \bar{l})$, termed Bijvoet pairs \cite{Bijvoet1954}, are not equivalent because of anomalous scattering.
The Flack parameter $x$~\cite{Flack1983, Berardinelli1985, Flack1999, Parson2013} is one of the fitting parameters in the structure analysis, and expressed as $I_{obs}(h,k,l)=(1-x)I_{cal}(h, k, l)+xI_{cal}(\bar{h}, \bar{k}, \bar{l})$, where $I_{obs}$ and $I_{cal}$ are the observed and calculated intensities, respectively.
The $I_{cal}(h,k,l)$ corresponds to the intensity calculated from the assumed chiral crystal structure in the analysis, while the $I_{cal}(\bar{h}, \bar{k}, \bar{l})$ is that for the opposite chiral structure.
When the deducted Flack parameter $x$ is refined to be zero, the assumed chiral crystal structure is correct.
When $x$ becomes one, the assumed chiral crystal structure has to be flipped into that with the opposite chirality.

Therefore, the handedness of the crystal structure was deducted by the Flack parameter $x$, which was obtained in the refined process of the analysis.
In the case of the right-handed crystal structure, $x$ becomes zero when the space group of the assumed structure is $P6_{2}22$.
In the case of the left-handed crystal structure, $x$ becomes zero for the assumed structure with the space group of $P6_{4}22$.

The results of the absolute structure analysis are summarized in Table \ref{tbl-NbTaSi2-conventional}.
The handedness of the crystals was left-handed for TaSi$_{2}$, while right-handed for NbSi$_{2}$.
These results are consistent with those reported in the literature \cite{Sakamoto2005}.
The lattice constants $a$ and $c$ of NbSi$_{2}$ are larger (below 0.5 percent) than those of TaSi$_{2}$. 
The lattice constants decrease with increasing the atomic number of $T$.
Namely, the values of $a$ and $c$ are proportional to the atomic radius of $T$ and can be used for determining $T$ atoms of the synthesized crystals.

\begin{table}[htbp]
      \caption{Crystal structure analysis of $T$Si$_{2}$, obtained by the LDFZ method using the feed rod made of the single composition (TaSi$_2$ for the rod \#1 and NbSi$_2$ for the rod \#2).}
      \label{tbl-NbTaSi2-conventional}

  \begin{center}
  
      \begin{tabular}{rrr}
      \hline
      \multicolumn{1}{c}{} & \multicolumn{1}{c}{TaSi$_{2}$} & \multicolumn{1}{c}{NbSi$_{2}$}\\
      \hline
      \multicolumn{1}{c}{Space group} & \multicolumn{1}{c}{$P6_{4}22$} & \multicolumn{1}{c}{$P6_{2}22$}\\
      \multicolumn{1}{c}{$a$ [\AA]} & \multicolumn{1}{c}{$4.7812(3)$} & \multicolumn{1}{c}{$4.8049(7)$}\\
      \multicolumn{1}{c}{$c$ [\AA]} & \multicolumn{1}{c}{$6.5672(4)$} & \multicolumn{1}{c}{$6.6034(5)$}\\
      \multicolumn{3}{c}{}\\
      \multicolumn{1}{c}{$R1$} & \multicolumn{1}{c}{$0.0241$} & \multicolumn{1}{c}{0.0255}\\
      \multicolumn{1}{c}{$wR2$} & \multicolumn{1}{c}{$0.0605$} & \multicolumn{1}{c}{0.0663}\\
      \multicolumn{1}{c}{Flack parameter $x$} & \multicolumn{1}{c}{$-0.076(175)$} & \multicolumn{1}{c}{$-0.109(89)$}\\
      \multicolumn{1}{c}{Absolute structure} & \multicolumn{1}{c}{Left-handed} & \multicolumn{1}{c}{Right-handed}\\
      \hline
      \end{tabular}

  \end {center}
\end{table}

Next, we performed the LDFZ crystal growth using the composition-gradient feed rod in order to obtain the single crystals of $T$Si$_{2}$ with the selected chirality, as shown in Fig.~\ref{f-FZ}(b).
The feed rod (\#3) was prepared for obtaining the left-handed NbSi$_{2}$ crystal.
It consists of TaSi$_{2}$ and NbSi$_{2}$ and thus the crystal growth starts with TaSi$_{2}$ in the region A, as shown in Fig.~\ref{f-FZ}(c).
The seed rod is initially crystallized into the single domain of TaSi$_{2}$. 
Then, the feed rod goes into the phase of NbSi$_{2}$ at the beginning of the region B.
In the successive crystal growth, NbSi$_{2}$ is crystallized on the seed crystal of TaSi$_{2}$, as shown in Fig.~\ref{f-FZ}(d).
If the crystallographic structure of NbSi$_{2}$ in the region B is inherited from that of TaSi$_{2}$ in the region A during the crystal growth, 
the grown NbSi$_{2}$ crystal should exhibit the chirality opposite to that of the crystal obtained by the conventional LDFZ method (shown in Table~\ref{tbl-NbTaSi2-conventional}).
The same technique was used to obtain the TaSi$_{2}$ crystals with the opposite handedness. In this case (\#4), the feed rod consists of NbSi$_{2}$ and TaSi$_{2}$ for the regions A and B, respectively.

Note that the melting temperature of TaSi$_{2}$ is 200 $^\circ$C higher than that of NbSi$_{2}$.
Thus, a drastic change of the melting temperature inevitably occurs at the boundary between the regions A and B.
In such a situation, it is difficult to stabilize the molten zone using conventional FZ furnaces with ellipsoid mirrors for focusing light emitted from halogen or xenon lamps. The temperature gradient of the conventional FZ furnaces is not as steep as that of the LDFZ furnaces. In this respect,
a precise control of the laser power can keep the molten zone stabilized in the LDFZ method using the well-focused laser.
This is one of the advantages of the crystal growth using the LDFZ method.

\begin{figure}[tb]
\begin{center}
\includegraphics[width=8cm]{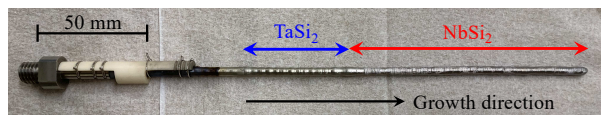}
\end{center}
\caption{An optical photograph of the single crystal, obtained by the LDFZ method with the composition-gradient feed rod that consists of TaSi$_{2}$ and NbSi$_{2}$.
}
\label{f-TSi-photo}
\end{figure}

Figure~\ref{f-TSi-photo} shows an optical photograph
of the single crystal obtained by using the composition-gradient feed rod (\#3). This rod crystal is made of TaSi$_{2}$ in the initial stage of the crystal growth and followed by the region of NbSi$_{2}$.
The crystalline orientation of the single crystal was evaluated by Laue X-ray backscattering method.
Both the TaSi$_{2}$ and NbSi$_{2}$ regions have the same crystallographic orientation, indicating that the crystallographic structure of TaSi$_{2}$ succeeds to that of NbSi$_{2}$.
The $c$-axis of the crystal is a few degrees tilted from the growth direction of the crystal.

In the case of crystals growth using the composition-gradient feed rod (\#4), the obtained rod kept the composition change from NbSi$_2$ to TaSi$_2$ along the direction of crystal growth as well.
The crystal of TaSi$_{2}$ obtained from the region B includes some grains with different crystallographic orientations.
This misorientation of the crystal growth may be caused by the sudden change of the melting temperature at the boundary between the regions A and B.
Note that the melting temperature in the region A is lower than that in the region B in the \#4 crystal growth.
In this case, when traveling the feed rod at the boundary, it was rather difficult to keep the molten zone stable because it became very slim due to a sudden solidification. More steps of the concentration from NbSi$_2$ to TaSi$_2$ are required in the composition distribution of the feed rod in order to make the melting point change gradually across the boundary.

To evaluate the crystal chirality of the grown rods (\#3 and \#4),
small pieces of the single crystals were prepared from both the TaSi$_{2}$ and NbSi$_{2}$ regions. 
The deducted Flack parameter $x$ was determined in such pieces by the absolute structure analysis using single crystal X-ray diffraction.
The results are summarized in Tables~\ref{tbl-TaNbSi2-gradient} and \ref{tbl-NbTaSi2-gradient}.
Note that the refined Flack parameters of all the pieces collected from the grown rods are the same within two-sigma error bar.

\begin{table}[htbp]
      \caption{Crystal structure analysis of $T$Si$_{2}$, obtained by the LDFZ method using the composition-gradient feed rod \#3.
                    It consists of TaSi$_{2}$ in the region A and NbSi$_{2}$ in the region B, as shown in Fig.~\ref{f-FZ}(a).}
      \label{tbl-TaNbSi2-gradient}

  \begin{center}

      \begin{tabular}{rrr}
      \hline
      \multicolumn{1}{c}{} & \multicolumn{1}{c}{TaSi$_{2}$} & \multicolumn{1}{c}{NbSi$_{2}$}\\
      \hline
      \multicolumn{1}{c}{Space group} & \multicolumn{1}{c}{$P6_{4}22$} & \multicolumn{1}{c}{$P6_{4}22$}\\
      \multicolumn{1}{c}{$a$ [\AA]} & \multicolumn{1}{c}{$4.7750(3)$} & \multicolumn{1}{c}{4.7904(2)}\\
      \multicolumn{1}{c}{$c$ [\AA]} & \multicolumn{1}{c}{$6.5550(5)$} & \multicolumn{1}{c}{6.5840(4)}\\
      \multicolumn{3}{c}{}\\
      \multicolumn{1}{c}{$R1$} & \multicolumn{1}{c}{$0.0115$} & \multicolumn{1}{c}{0.0074}\\
      \multicolumn{1}{c}{$wR2$} & \multicolumn{1}{c}{$0.0249$} & \multicolumn{1}{c}{0.0164}\\
      \multicolumn{1}{c}{Flack parameter $x$} & \multicolumn{1}{c}{$0.057(136)$} & \multicolumn{1}{c}{$0.021(148)$}\\
      \multicolumn{1}{c}{Absolute structure} & \multicolumn{1}{c}{Left-handed} & \multicolumn{1}{c}{Left-handed}\\
      \hline
      \end{tabular}

  \end {center}
\end{table}

\begin{table}[htbp]
      \caption{Crystal structure analysis of $T$Si$_{2}$, obtained by the LDFZ method with the composition-gradient feed rod \#4.
                    It consists of NbSi$_{2}$ in the region A and TaSi$_{2}$ in the region B as shown in Fig.~\ref{f-FZ}(a).}
      \label{tbl-NbTaSi2-gradient}

  \begin{center}

      \begin{tabular}{rrr}
      \hline
      \multicolumn{1}{c}{} & \multicolumn{1}{c}{NbSi$_{2}$} & \multicolumn{1}{c}{TaSi$_{2}$}\\
      \hline
      \multicolumn{1}{c}{Space group} & \multicolumn{1}{c}{$P6_{2}22$} & \multicolumn{1}{c}{$P6_{2}22$}\\
      \multicolumn{1}{c}{$a$ [\AA]} & \multicolumn{1}{c}{$4.7975(2)$} & \multicolumn{1}{c}{4.7770(4)}\\
      \multicolumn{1}{c}{$c$ [\AA]} & \multicolumn{1}{c}{$6.5837(4)$} & \multicolumn{1}{c}{6.5604(5)}\\
      \multicolumn{3}{c}{}\\
      \multicolumn{1}{c}{$R1$} & \multicolumn{1}{c}{$0.0164$} & \multicolumn{1}{c}{0.0287}\\
      \multicolumn{1}{c}{$wR2$} & \multicolumn{1}{c}{$0.0371$} & \multicolumn{1}{c}{0.0745}\\
      \multicolumn{1}{c}{Flack parameter $x$} & \multicolumn{1}{c}{$-0.077(145)$} & \multicolumn{1}{c}{$-0.021(153)$}\\
      \multicolumn{1}{c}{Absolute structure} & \multicolumn{1}{c}{Right-handed} & \multicolumn{1}{c}{Right-handed}\\
      \hline
      \end{tabular}

  \end {center}
\end{table}

For the crystal growth using the feed rod (\#3), the Flack parameter $x$ of TaSi$_{2}$ in the region A is almost zero with regard to the structure based on the space group of $P6_{4}22$, as shown in Table~\ref{tbl-TaNbSi2-gradient}.
Thus, the obtained crystal of TaSi$_{2}$ has the left-handed crystal structure.
Importantly, the crystal handedness of NbSi$_{2}$ turns to be the same as that of TaSi$_{2}$, while the lattice constants in the NbSi$_{2}$ region reproduces the value obtained by the conventional FZ method. 

The handedness control was also performed by using the feed rod (\#4) shown in Table~\ref{tbl-NbTaSi2-gradient}.
For the grown rod, the crystal handedness of NbSi$_{2}$ in the region A and that of TaSi$_{2}$ in the region B were determined to be right-handed.
As for the uniformity of chirality of TaSi$_{2}$ in the region B, it turned out that all the grains examined, including the ones with different crystallographic orientations from that of NbSi$_{2}$ in the region A, exhibit the right-handed crystal structure. In this respect, the crystal handedness of the region B is likely to inherit from that of the region A even though some misorientation occurs during the crystal growth.

These experiments demonstrate that the single crystals of left-handed NbSi$_2$ and right-handed TaSi$_2$ were successfully obtained. These crystals have the handedness opposite to that of the crystals grown by the conventional methods.
Once the single crystals with the opposite chirality are obtained, such crystals can be used as a seed crystal for another crystal growth. 
For instance, by feeding polycrystalline NbSi$_{2}$ on the chirality-controlled seed rod, the NbSi$_{2}$ crystals with the opposite chirality are successively obtained without using the composition-gradient feed rod.
This method can be applicable for controlling chirality of $T$Si$_{2}$ with different $T$ element.
We emphasize that a preparation of the chirality-controlled seed crystals is essential in the chirality-controlled crystal growth.

\subsection{CISS measurements}

\begin{figure}[tb]
\begin{center}
\includegraphics[width=8cm]{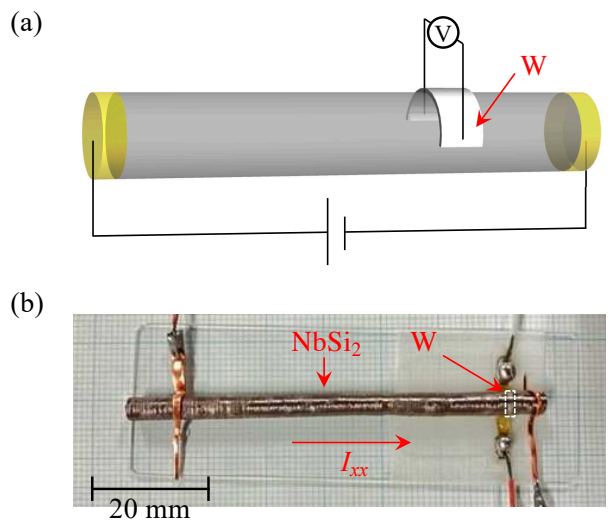}
\end{center}
\caption{(a) Schematics of the CISS measurements for a cylinder-shaped specimen. (b) A photograph of the device made for the CISS measurements of the NbSi$_{2}$ crystal with the tungsten (W) electrode. The $I$-$V$ characteristics were collected by applying the charge current into the crystal.
}
\label{f-CISS-draw}
\end{figure}

The CISS measurements were performed with the chiral crystal rods, grown by the LDFZ method, as illustrated in Fig.~\ref{f-CISS-draw}(a).
The spin polarization is monitored by a transverse voltage generated in a tungsten (W) electrode, deposited on a half-cylindrical region of the crystal.

For the CISS measurements, we prepared the chirality-controlled NbSi$_2$ crystals of a cylinder shape with a  dimension of 5 mm in diameter and 70 mm in length, as shown in Fig.~\ref{f-CISS-draw}(b).
More accurately, the LDFZ grown crystals with the right- and left-handedness were cut into a slice with 70 mm long.
The W electrode with 1.5 mm in width and 7 nm in thickness was deposited on the specimen by using a sputtering system (M09-0014, SEED Lab. Corporation).
The electrodes for the current injection are set to be 60 mm apart from each other and contain the W electrode between them.   

\begin{figure}[tb]
\begin{center}
\includegraphics[width=8cm]{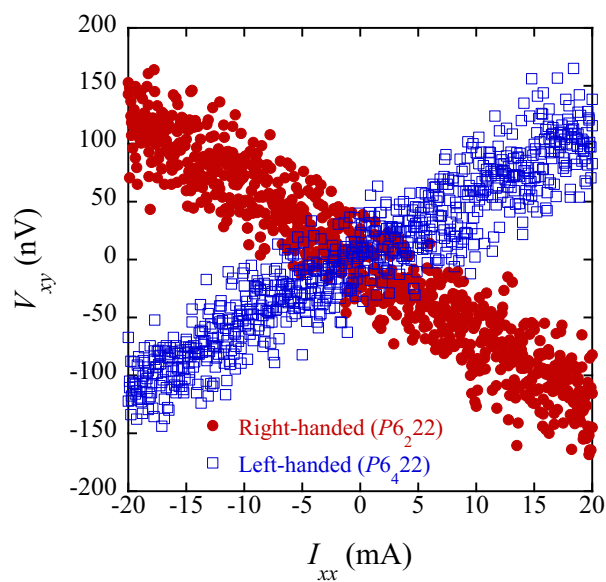}
\end{center}
\caption{
The $I$-$V$ characteristics of NbSi$_{2}$ collected at room temperature without magnetic field.
The CISS voltage $V_{xy}$ was measured as a function of $I_{xx}$ in the right-handed ($P$6$_{2}$22) and left-handed ($P$6$_{4}$22) NbSi$_{2}$ crystals with the tungsten electrodes.
}
\label{f-CISS}
\end{figure}

The longitudinal electrical current $I_{xx}$ was applied along the long axis of the crystal. 
As evaluated by Laue X-ray, the longitudinal direction of the crystals is deviated from the $c$-axis by several degrees and thus the $I_{xx}$ flows in a direction nearly parallel to the $c$-axis.

In a separate experiment, the electrical resistivity $V_{xx}/I_{xx}$ was measured. It was determined to be $39\mu\Omega\cdot$cm at room temperature, which was slightly smaller than those of single crystalline NbSi$_{2}$ obtained by the Czochralski method \cite{Onuki2014}.
This value was an order of magnitude smaller than that of polycrystalline NbSi$_{2}$ \cite{Shishido2021}.

The CISS signals were collected with the NbSi$_{2}$ crystals at room temperature without magnetic field. 
The transverse voltage $V_{xy}$ was generated linearly as a function of $I_{xx}$ in the NbSi$_{2}$ crystals, as shown in Fig.~\ref{f-CISS}.

Importantly, the sign of the linear $I$-$V$ characteristics depends on the handedness of the crystals. 
The sign reversal occurs because of the spin polarization flipping accompanied by a change of the crystal chirality.
These results indicate that the CISS measurements can probe the handedness of the crystal over centimeters and work as a macroscopic method for chirality determination.  

Note that the $V_{xy}$ signals mainly result from the CISS effect rather than longitudinal electrical resistance of NbSi$_{2}$ caused by a misalignment of the W electrodes.
If the signals completely come from the misalignment, they correspond to a resistance of the specimen of 0.3 mm in length.
Taking the width of the W electrode (1.5 mm) into account, it is hard to see such a large misalignment of the electrode deposited on the rod.

Note that the magnitude of the slope $V_{xy} / I_{xx}$ in the present study is three orders magnitude smaller than that obtained in the micrometer-sized single crystals \cite{Shiota2021},
while it is an order of magnitude smaller than that for polycrystals \cite{Shishido2021}.
However, by converting $V_{xy} / I_{xx}$ into the resistivity $\rho_{xy}$ using the specimen dimensions and then normalized by $\rho_{xx}$, the resultant $\rho_{xy} / \rho_{xx}$ exhibits the same order of the magnitude as those for the micrometer-sized single crystals and for the polycrystals.

The spatial-resolved CISS measurements revealed the presence of chirality domains with the opposite chirality in the micrometer-sized specimen \cite{Shiota2021}.
The bulk crystals used in the present study may also contain such chirality domains. 
The Flack parameter determined by the single crystal X-ray diffraction measurements indicates the presence of such chirality domains up to a few percent of all the region within its error bar. 
However, there is no direct information on a microscopic distribution of the chirality domains in the bulk crystal. 
A further quantitative discussion requires for identifying a locational distribution of the chirality domains using the spatial-resolved CISS measurements in a resolution of micrometers or smaller scale.
Such precise CISS measurements for the centimeter-sized crystals will be a future issue.

\section{Summary and Perspective}
In summary, we grew the chirality-selected single crystals of NbSi$_{2}$ and TaSi$_{2}$ using the LDFZ method.
In the case of the conventional LDFZ growth using the seed and feed rods that consist of the same compound, the right-handed NbSi$_{2}$ the left-handed TaSi$_{2}$ crystals were obtained as reported in the literature.
However, in the case of the LDFZ crystal growth using the composition-gradient feed rods,
the chirality of the obtained NbSi$_{2}$ crystals was found to be the left-handed,
while that of the TaSi$_{2}$ crystals was the right-handed.
The chirality is opposite to the reported one in both crystals
and is hardly obtained by the conventional crystal growth technique.
When such crystals with the opposite chirality are used for the seed crystals, 
the chiral crystal of the same quality in terms of chirality can be obtained even by using the conventional LDFZ method.
These results indicate that the chirality of the seed crystal is transferred to the grown crystal. 

The spin polarization phenomena were observed in the rod-shaped single crystals of NbSi$_{2}$ with millimeters in diameter and centimeters in length.
The sign of the slope in the $I$-$V$ characteristics for the NbSi$_{2}$ bulk rod crystals was qualitatively consistent with the results obtained with the micrometer-sized specimens of NbSi$_{2}$.
The CISS effect appears in such large crystals, and will open up new possibilities for a non-destructive evaluation of chiral crystal structures in inorganic materials.

The next interesting issue will be an investigation of the locational distribution of the crystal chirality over the long crystal.
The distribution of the chirality can also be detected by a circular-polarized resonant X-ray micro diffraction method \cite{Ohsumi2013}.
However, this method proves only on the surface of the specimen because of very large X-ray absorption.
On the other hand, the CISS signal reflects the crystal chirality in the specimen where the current is injected \cite{Inui2020,Shiota2021,Shishido2021}.
When the current is injected in a very narrow region of the specimen, the chirality distribution can be evaluated with a high spatial resolution.

% acknowledgments
\ack

This work was supported by JSPS KAKENHI Grant Numbers 15H05885, 19KK0070 and 20H02642.

\end{document}